\documentclass[10pt,conference]{IEEEtran}
\usepackage{amsmath}
\newtheorem{theorem}{Theorem}[section]

\newtheorem{definition}[theorem]{Definition}
\newtheorem{lemma}[theorem]{Lemma}
\newtheorem{example}[theorem]{Example}
\newtheorem{corollary}[theorem]{Corollary}
\newtheorem{remark}[theorem]{Remark}

\font\msbm=msbm10 at 10pt

\newcommand{\FF}{\mbox{\msbm F}}

\newcommand{\CC}{\mbox{\msbm C}}

\begin{document}

\title{Grassmannian Codes as Lifts of Matrix Codes Derived as Images of Linear Block Codes \\over Finite Fields}
\author{
   \IEEEauthorblockN{Bryan S. Hernandez and Virgilio P. Sison}
   \IEEEauthorblockA{Institute of Mathematical Sciences and Physics\\
     University of the Philippines, Los Ba\~{n}os\\
     College, Laguna 4031, Philippines\\
     Email: \{bshernandez, vpsison\}@up.edu.ph} 
 }
\maketitle
\begin{abstract}
\boldmath Let $p$ be a prime such that $p \equiv 2$ or 3 (mod 5). Linear block codes over the non-commutative matrix ring $M_2(\FF_p)$ endowed with the Bachoc weight are derived as isometric images of linear block codes over the Galois field $\FF_{p^2}$ endowed with the Hamming metric. When seen as rank metric codes, this family of matrix codes satisfies the Singleton bound and thus are maximum rank distance codes, which are then lifted to form a special class of subspace codes, the Grassmannian codes, that meet the anticode bound. These so-called anticode-optimal Grassmannian codes are associated in some way with complete graphs. New examples of these maximum rank distance codes and anticode-optimal Grassmannian codes are given.
\end{abstract}
\vskip .1in

\section{Introduction}
\label{sec:int}
This paper deals with certain concepts of ``coding theory in projective space" and highlights the practical significance of subspace codes, specifically of Grassmannian codes, in error correction in networks. Let $q=p^r$, $p$ a prime, $r$ a positive integer, and $\FF_q$ the Galois field with cardinality $q$ and characteristic $p$. Consider the $n$-dimensional full vector space $\FF_q^n$ over $\FF_q$. The set of all subspaces of $\FF_q^n$, denoted by $\mathcal P_q(n)$, is called the projective space of order $n$ over $\FF_q$.   For an integer $k$, where $0 \le k \le n$, the set of all $k$-dimensional subspaces of $\FF_q^n$, denoted by $\mathcal G_q(n,k)$, is called the Grassmannian. A subspace code is a nonempty subset of $\mathcal P_q(n)$. A Grassmannian code is a nonempty subset of $\mathcal G_q(n,k)$ which is also called a constant dimension code, that is, the codewords in $\mathcal G_q(n,k)$ are subspaces of $\FF_q^n$ of dimension $k$, thus they are nothing but rate-$k/n$ linear block codes of length $n$ over $\FF_q$. Subspace codes have practical importance in network coding. The seminal paper \cite{ahlswede} refers to \textit{network coding} as ``coding at a node in a network", that is, a node receives information from all input links, then encodes and sends information to all output links.  

Section II gives important theoretical preliminaries, while Section~\ref{grassmannian} shows how to construct Grassmannian codes endowed with the subspace distance as lifts of certain linear block codes $\mathcal M$ over the non-commutative matrix ring $M_2(\FF_p)$ endowed with the Bachoc weight. The matrix codes  $\mathcal M$ are isometric images of linear block codes over $\FF_{p^2}$ endowed with the Hamming distance. More importantly, these matrix codes are maximum rank distance codes, or MRD codes, that is, they satisfy the Singleton bound for matrix codes with respect to the rank metric. The Grassmannian codes constructed from these lifts are anticode-optimal, or simply optimal, in the sense that they satisfy the anticode bound. New examples of MRD codes and anticode-optimal Grassmannian codes are given from these constructions. In Section~\ref{complete:graphs} it is shown that this family of anticode-optimal Grassmannian codes can be associated in a peculiar way with complete graphs. 

\section{Preliminaries}
\label{prelim}
The set of all $k\times\ell$ matrices over $\FF_q$, denoted by $M_{k\times\ell}(\FF_q)$, is considered as a vector space over $\FF_q$. A nonempty subset of $M_{k \times \ell}(\FF_q)$ is called a $[k \times \ell]$ {\it matrix code} over $\FF_q$. This $[k\times\ell]$ matrix code is said to be linear if it is a subspace of $M_{k \times \ell}(\FF_q)$.

The {\it rank distance} between two $k \times \ell$ matrices over $\FF_q$, say $A$ and $B$, is defined by $d_R(A,B)= $ rank$(A-B)$, and is clearly a metric. A $[k \times \ell,\delta]$ {\it rank-metric code} $\CC$ is a $[k \times \ell]$ matrix code whose minimum rank distance is $\delta$. That is, $\delta=\min\{d_R(A,B)|A,B \in \CC, A\neq B\}$. 

\begin{definition}
A $[k\times \ell,\rho,\delta]$ {\it rank-metric code} is a linear code in $M_{k \times \ell} (\FF_q)$ with dimension $\rho$ and minimum rank distance $\delta$. 
\end{definition}

The following theorem gives the Singleton bound for rank-metric codes.

\begin{theorem}{(T. Etzion and A. Vardy, \cite{etz})}
\label{singleton}
For a $[k\times \ell,\rho,\delta]$ rank-metric code, we have $\rho \leq \min\{k(\ell-\delta+1),\ell(k-\delta+1)\}.$
\end{theorem}

A code that attains this bound is called a {\it{maximum rank distance code}} or an {\it{MRD code}}. The only previously known examples of MRD codes are the so-called Gabidulin codes. 

\begin{definition}
Let $A \in M_{k \times \ell} (\FF_q)$. The {\it{lift}} of $A$, denoted by $L(A)$, is the $k\times(k+\ell)$ standard matrix $(I_k$ $A)$, where $I_k$ is the $k \times k$ identity matrix.
\end{definition}

The subspace generated by the rows of the lifted matrix $L(A)$ will be denoted by $\langle L(A) \rangle$. This subspace is in fact a rate-$k/(k+\ell)$ linear block code of length $k + \ell$ over $\FF_q$.

There are at least two metrics that can be applied on the projective space $\mathcal P_q(n)$. The {\it{subspace distance}}  is given by $ d_S(A,B)=\dim A + \dim B - 2\dim(A \cap B)$. The next one is the {\it{injection distance}} which is given by $d_I(A,B)=\max\{\dim A ,\dim B\} - \dim(A \cap B)$, for all $A,B \in \mathcal P_q(n)$. In this paper we shall use the subspace distance on the constructed Grassmannian codes.

A classic formula for the cardinality of the Grassmannian $\mathcal G _q(n,k)$ is given by the $q$-ary Gaussian coefficient $$ {n \brack k}_q = \prod^{k-1}_{i=0} \frac{q^n-q^i}{q^k-q^i}.$$ 

\begin{definition}
A Grassmannian code $\mathcal C$ in $\mathcal G _q (n,k)$ is called an $(n,M,d,k)_q$ code if $|\mathcal C|=M$ and its minimum subspace distance is $d$, where $d=\{\min d_S(U,V)|U,V \in \mathcal C, U\neq V\}.$ 
\end{definition}

\begin{definition}
Let $\mathcal C$ be a $[k \times \ell]$ rank-metric code. The set 
\begin{align*}
\Lambda (\mathcal C)&=\{\langle L(A)  \rangle|A \in \mathcal C \}\\
&=\{\langle (I_k \  A)  \rangle|A \in \mathcal C \}
\end{align*}
is called the {\it{lift}} of $\mathcal C$, which is clearly a Grassmannian code. The next theorem gives a more specific result.
\end{definition}

\begin{theorem}{(A. Khaleghi and D. Silva and F. R. Kschischang, \cite{khaleghi})}
\label{rank:metric}
Let $\mathcal C$ be a $[k \times \ell,\rho,\delta]$ rank-metric code. The lift of  $\mathcal C$ is a $(k+\ell, q^\rho,2\delta, k)_q$ Grassmannian code.
\end{theorem}

The next results give bounds on the maximum number of codewords in a Grassmannian code.
\begin{theorem}{(T. Etzion and A. Vardy, \cite{etzion:vardy})}
Let $\mathcal A_q(n,d,k)$ be the maximum number of codewords of a code in $\mathcal G_q(n,k)$ with subspace distance $d=2\delta+2$. Then 
\[\mathcal A_q(n,2\delta+2,k) \le \frac{{n \brack {k-\delta}}_q}{{ k \brack {k-\delta}}_q  }.\]
\label{antibound1}
\end{theorem}

\begin{theorem}(A. Khaleghi, D. Silva, F. R. Kschischang, \cite{khaleghi})
Let $\mathcal A_q(n,d,k)$ be the maximum number of codewords of a code in $\mathcal G_q(n,k)$ with injection distance $d$. Then 
\[\mathcal A_q(n,d,k) \le \frac{{n \brack {k-d+1}}_q}{{ k \brack {k-d+1}}_q  }.\]
\label{antibound2}
\end{theorem}

Theorems \ref{antibound1} and \ref{antibound2} are two existing versions of the {\it{Anticode Bound}} for Grassmannian codes.

If $A \in \FF_q^n$, the {\it dual or orthogonal subspace} of $A$ is given by $A^\perp=\{v \in V|u\cdot v =0$ for all $u\in A\}$ where $u\cdot v$ is the usual inner product between vectors $u$ and $v$. As in the classical sense, there is the notion of duality in Grassmannian codes.

\begin{definition}{(R. K\"otter and F. R. Kschichang, \cite{kk})}

If $\mathcal C \subseteq \mathcal G_q(n,k)$ then its dual or complementary code is given by $\mathcal C^{\perp}=\{C^{\perp} \in \mathcal G_q(n,n-k) | C \in \mathcal C\}.$

\end{definition}

\begin{theorem}{(R. K\"otter and F. R. Kschichang, \cite{kk})}
If $\mathcal C$ is an $(n, M, d, k)_q$ code then $\mathcal C^{\perp}$ is an $(n, M, d, n-k)_q$ code.
\label{ortho}
\end{theorem}
\section{Rank-Metric Codes and Grassmannian Codes }
\label{grassmannian}
Let $M_2(\FF_p)$ be the non-commutative ring of $2 \times 2$ matrices over $\FF_p$ and $GL(2,p)$ its multiplicative group of units. We now give the definition of the Bachoc weight $w_{\tt B}$ on $M_2(\FF_p)$. .
\begin{gather*}
		   w_{\tt B}(A) = \begin{cases}
       		 0 & {\rm if} \ A = { 0} \\
                         1 & {\rm if} \ A\in GL(2,p)\\
     		 p & {\text{otherwise}}\
      		  \end{cases}
		\end{gather*}
In \cite{bachoc}, an isometric map $\phi$ from $\FF_4^2$ onto $M_2(\FF_2)$ where \[ \phi((a+b \omega, c+d \omega))={\begin{pmatrix}
                                                      a+d & b+c  \\
                                                      b+c+d & a+b+d 
                                           \end{pmatrix}}\]
was given using the Hamming weight $w_{\tt Ham}$ and the Bachoc weight $w_{\tt B}$ for $\FF_4^2$ and $M_2(\FF_2)$ respectively, such that $w_{\tt Ham}(\alpha)=w_{\tt B}(\phi(\alpha))$ for all $\alpha \in \FF_4^2$.

Table \ref{bachoc_rank} shows the elements of $\FF_4^2$ with their corresponding Hamming weights and the elements of $M_2(\FF_2)$ with their corresponding Bachoc and rank weights.
\begin{table}
\centering
\begin{tabular}{|c|c|c|c|c|}
\hline
   $\alpha$ & $ w_{\tt Ham}(\alpha)$
& $\phi(\alpha)$ & $w_{\tt B}(\phi(\alpha))$ & $w_R(\phi(\alpha))$\\
\hline
$(0,0)$ & 0 & ${\begin{pmatrix}
                                                      0 & 0  \\
                                                      0 & 0 
                                           \end{pmatrix}}$ & 0 & 0  \\
\hline

$(0,1)$ & 1 & ${\begin{pmatrix}
                                                      0 & 1  \\
                                                      1 & 0 
                                           \end{pmatrix}}$ & 1 & 2  \\
\hline

$(1,0)$ & 1 & ${\begin{pmatrix}
                                                      1 & 0  \\
                                                      0 & 1 
                                           \end{pmatrix}}$ & 1 & 2  \\
\hline

$(1,1)$ & 2 & ${\begin{pmatrix}
                                                      1 & 1  \\
                                                      1 & 1 
                                           \end{pmatrix}}$ & 2 & 1  \\
\hline

$(0,\omega)$ & 1 & ${\begin{pmatrix}
                                                      1 & 0  \\
                                                      1 & 1 
                                           \end{pmatrix}}$ & 1 & 2  \\
\hline

$(\omega,0)$ & 1 & ${\begin{pmatrix}
                                                      0 & 1  \\
                                                      1 & 1 
                                           \end{pmatrix}}$ & 1 & 2  \\
\hline

$(\omega,\omega)$ & 2 & ${\begin{pmatrix}
                                                      1 & 1  \\
                                                      0 & 0 
                                           \end{pmatrix}}$ & 2 & 1  \\
\hline

$(1,\omega)$ & 2 & ${\begin{pmatrix}
                                                      0 & 0  \\
                                                      1 & 0 
                                           \end{pmatrix}}$ & 2 & 1  \\
\hline

$(\omega,1)$ & 2 & ${\begin{pmatrix}
                                                      0 & 0  \\
                                                      0 & 1 
                                           \end{pmatrix}}$ & 2 & 1  \\
\hline

$(0,1+\omega)$ & 1 & ${\begin{pmatrix}
                                                      1 & 1  \\
                                                      0 & 1 
                                           \end{pmatrix}}$ & 1 & 2  \\
\hline

$(1+\omega,0)$ & 1 & ${\begin{pmatrix}
                                                      1 & 1  \\
                                                      1 & 0 
                                           \end{pmatrix}}$ & 1 & 2  \\
\hline

$(1,1+\omega)$ & 2 & ${\begin{pmatrix}
                                                      0 & 1  \\
                                                      0 & 0 
                                           \end{pmatrix}}$ & 2 & 1  \\
\hline

$(1+\omega,1)$ & 2 & ${\begin{pmatrix}
                                                      1 & 0  \\
                                                      0 & 0 
                                           \end{pmatrix}}$ & 2 & 1  \\
\hline

$(\omega,1+\omega)$ & 2 & ${\begin{pmatrix}
                                                      1 & 0  \\
                                                      1 & 0 
                                           \end{pmatrix}}$ & 2 & 1  \\
\hline

$(1+\omega,\omega)$ & 2 & ${\begin{pmatrix}
                                                      0 & 1  \\
                                                      0 & 1 
                                           \end{pmatrix}}$ & 2 & 1  \\
\hline

$(1+\omega,1+\omega)$ & 2 & ${\begin{pmatrix}
                                                      0 & 0  \\
                                                      1 & 1 
                                           \end{pmatrix}}$ & 2 & 1  \\
\hline
\end{tabular}
\caption{Hamming Weights on $\FF_4^2$ and Bachoc and Rank Weights on $M_2(\FF_2)$}
\label{bachoc_rank}
\end{table}

\begin{lemma}
\label{minrank}
Let $\mathcal C$ be a $[k \times \ell, \rho, \delta]$ rank-metric code with minimum nonzero rank $\Omega$. Then $\delta=\Omega$.
\end{lemma}

\begin{IEEEproof} 
Let $\mathcal C$ be a rank-metric code with minimum rank distance $\delta$ and minimum nonzero rank $\Omega$. Let $A$ and $B$ be distinct elements of $\mathcal C$ such that ${\text{rank}}(A-B)$ is minimum. Note that ${\text{rank}}(A-B) \ne 0$.  Then $\delta=d_R(A,B)={\text{rank}}(A-B) \ge \Omega$. Moreover, let $A \in \mathcal C$ with minimum rank. Now, $\Omega={\text{rank}}(A)=d_R(A,0) \ge \delta$. Thus, $\delta = \Omega$.
\end{IEEEproof} 

\begin{lemma}
Let $q$ be a power of a prime $p$. Then $\FF_q^n$ is also an $\FF_p$-vector space.
\label{fqnvec}
\end{lemma}

\begin{lemma}
\label{propphi}
Let $\phi: {\left(\FF_{p^2}\right)}^2 \longrightarrow M_2(\FF_p)$ where \[ \phi((a+b \omega, c+d \omega))={\begin{pmatrix}
                                                      a+d & b+c  \\
                                                      b+c+d & a+b+d 
                                           \end{pmatrix}}.\]
Then $\phi$ is an isomorphism of $\FF_p$-vector spaces.
\end{lemma}

\begin{remark}
\label{isovector}
From Lemma \ref{propphi}, if $C$ is a linear block code of length 2 over $\FF_{p^2}$ then $C \cong \phi  (C)$ as $\FF_p$-vector spaces.
\end{remark}

For the following remark, let $\alpha_i=(a_i+b_i \omega, c_i+d_i \omega)\in \left(\FF_{p^2}\right)^2$ and 
\[A_i= {\begin{pmatrix}
                                                      a_i+d_i & b_i+c_i  \\
                                                     b_i+c_i+d_i & a_i+b_i+d_i
                                           \end{pmatrix}}\in M_{2}(\FF_p)\]
where $1 \le i \le r$ for positive integer $r$.

\begin{remark}
Let $r$ be a positive integer. Note that $\phi$ can be extended naturally in the following manner. We have \\$\phi: {\left(\FF_{p^2}\right)}^{2r} \longrightarrow M_{2 \times 2r}(\FF_p)$ where \[ \phi(\alpha_1,\alpha_2,...,\alpha_{2r})=
{\begin{pmatrix}
                                                      A_1    &      A_2  & ...    &   A_r \end{pmatrix}}.\]
It is easy to see that ${\left(\FF_{p^2}\right)}^{2r} \cong M_{2 \times 2r}(\FF_p)$ as $\FF_p$-vector spaces.
If $C$ is a linear block code of length $2r$ over $\FF_{p^2}$ then $C \cong \phi (C)$ as $\FF_p$-vector spaces.
\label{isovector2}
\end{remark}

\begin{lemma}{(D. Falcunit, Jr. and V. Sison, \cite{falcunit:sison})}
\label{irreducible}
If $p \equiv$ $2$ or $3$ (mod $5$) then the polynomial $f(x)=x^2+x+(p-1)$ is irreducible over $\FF_{p}$.
\end{lemma}

\begin{theorem}
Let $C$ be a linear block code of length $2r$ over $\FF_{p^2}$ and $\rho$ its dimension as an $\FF_{p^2}$-vector space. 
If $p \equiv 2$ or 3 (mod 5) and for all $(\alpha _1, \alpha _2, ... ,\alpha _{2r}) \in C$, $\alpha_j=0$ for each odd (resp. even) index $j$, then
\begin{enumerate}
\item[i.] $\phi(C)$ is a $[2 \times 2r, \rho, 2]$ rank-metric code,
\item[ii.] $\Lambda(\phi (C))$ is a $(2r+2,p^\rho,4,2)_p$ code, and;
\item[iii.] the pairwise intersection of codewords of $\Lambda(\phi (C))$ is trivial.
\end{enumerate}
\label{classtonew2}
\end{theorem}

\begin{IEEEproof}
Let $C$ be a linear block code of length $n$ over $\FF_{p^2}$.
Note that by Remark \ref{isovector2}, $C$ and $\phi (C)$ are isomorphic as $\FF_p$-vector spaces. Hence, the dimension of $\phi (C)$ is $\rho$.
Moreover, let $(\alpha _1, \alpha _2, ... ,\alpha _{2r}) \in C \backslash \{(0,0, ..., 0)\}$, $\alpha_j=0$ for each odd (resp. even) integer $j$.
To simplify the proof, we consider when $r=1$ and hence we have $(0, \alpha _2) \in C \backslash \{(0,0)\}$. Note that $\alpha _2=c+d\omega$ for some $c,d \in \FF_p$. Then $\phi(0,c+d\omega)={\begin{pmatrix}
                                                      d & c  \\
                                                     c+d & d
                                           \end{pmatrix}}$.
Since $c$ and $d$ are not both zero, we have the following cases:
\begin{enumerate}
\item[1.] If $c=0$ and $d\ne 0$ then the matrix becomes
${\begin{pmatrix}
                                                      d & 0  \\
                                                      d & d
                                           \end{pmatrix}}$ with rank 2.
\item[2.] If $c \ne 0$ and $d=0$ then the matrix becomes
${\begin{pmatrix}
                                                      0 & c  \\
                                                      c & 0
                                           \end{pmatrix}}$ with rank 2.
\item[3.] Let $c,d \ne 0$. Suppose rank of the matrix is not 2 then one row is a multiple of the other, that is, $(d,c)=x(c+d,d)$ for some $x \in \FF_p$. This implies that $d=xc+xd$ and $c=xd$. Further, $d=x^2d+xd$ and $x^2+x-1=0$. Since $p \equiv 2$ or 3 (mod 5), by Lemma \ref{irreducible}, $f(x)=x^2+x-1=x^2+x+(p-1)$ is irreducible over $\FF_p$. Thus there is no $x \in \FF_p$ such that $(d,c)=x(c+d,d)$ and hence the rank of the matrix is 2.
\end{enumerate}
Thus, the minimum rank weight of $\phi (C)$ is 2. By Lemma \ref{minrank}, the minimum rank distance of $\phi (C)$ is also 2. It follows that $\phi (C)$ is a $[2 \times 2r, \rho, 2]$ rank-metric code.

It easy to see that (ii) follows directly from Theorem \ref{rank:metric}.

If $\Lambda(\phi (C))$ is a $(2r+2,p^{\rho},4,2)_2$ code, the minimum subspace distance of $\Lambda(\phi (C))$ is 4. Let $A, B \in \Lambda(\phi (C))$. Note that $\dim A=\dim B=2$ and we have $4 \le d_S(A,B)=\dim A +\dim B - 2\dim (A \cap B)$. Thus, $4 \le 2+2-2\dim(A \cap B)$ and hence $\dim (A \cap B) \le 0$. Therefore, $\dim (A \cap B)=0$.
This means that the pairwise intersection of codewords of $\Lambda(\phi (C))$ is trivial. \end{IEEEproof} 

\begin{remark}
Let $r$ be a positive integer and consider \[S=\{(0,c_1+d_1 \omega,0,c_2+d_2 \omega,0,...,0,c_r+d_r \omega)|c_i,d_i \in \FF_p\},\] a subspace of $\left(\FF_{p^2}\right)^{2r}$ as an $\FF_p$-vector space. By Theorem \ref{classtonew2}, we can consider the map $\phi: S \longrightarrow M_{2 \times 2r}(\FF_p)$ where \[ \phi((0, c_1+d_1 \omega, 0, c_2+d_2 \omega,..., 0, c_{r}+d_{r} \omega)) =\] \[{\begin{pmatrix}
                                                      d_1 & c_1 &  d_2 & c_2 & ...  & d_{r} & c_{r}  \\
                                                      c_1+d_1 & d_1  & c_2+d_2 & d_2 & ... & c_{r}+d_{r} & d_{r}
                                           \end{pmatrix}}\]
as the map 
$\phi _O: \left(\FF_{p^2}\right)^r \longrightarrow M_{2 \times 2r}(\FF_p)$ where \[ \phi_O((c_1+d_1 \omega, c_2+d_2 \omega,..., c_{r}+d_{r} \omega))\] \[={\begin{pmatrix}
                                                      d_1 & c_1 &  d_2 & c_2 & ...  & d_{r} & c_{r}  \\
                                                      c_1+d_1 & d_1  & c_2+d_2 & d_2 & ... & c_{r}+d_{r} & d_{r}
                                           \end{pmatrix}}.\]
Note that $\phi(S)=\phi _O\left(\left(\FF_{p^2}\right)^r\right)$ and hence the rank of each nonzero element of $\phi _O\left(\left(\FF_{p^2}\right)^r\right)$ is 2. Since the odd positions of $S$ are all zeros, we can collapse the elements in such a way that the entries in the odd positions are deleted and hence we can look at the elements of $S$ as simply elements of $\left(\FF_{p^2}\right)^r$.

In an analogous manner consider \[\overline{S}=\{(a_1+b_1 \omega,0,a_2+b_2 \omega,0,...,a_r+b_r \omega,0)|a_i,b_i \in \FF_p\},\] which is also a subspace of $\left(\FF_{p^2}\right)^{2r}$ as an $\FF_p$-vector space. By Theorem \ref{classtonew2} iv, we can look at $\phi: \overline{S} \longrightarrow M_{2 \times 2r}(\FF_p)$ where \[ \phi((a_1+b_1 \omega,0,a_2+b_2 \omega,0,...,a_r+b_r \omega,0))\] \[={\begin{pmatrix}
                                                      a_1 & b_1 &  a_2 & b_2 & ...  & a_{r} & b_{r}  \\
                                                      b_1 & a_1+b_1  & b_2 & a_2+b_2 & ... & b_{r} & a_{r}+b_{r}
                                           \end{pmatrix}}\]
as
$\phi _E: \left(\FF_{p^2}\right)^r \longrightarrow M_{2 \times {2r}}(\FF_p)$ where \[ \phi _E((a_1+b_1 \omega, a_2+b_2 \omega,..., a_{r}+b_{r} \omega))\] \[={\begin{pmatrix}
                                                       a_1 & b_1 &  a_2 & b_2 & ...  & a_{r} & b_{r}  \\
                                                      b_1 & a_1+b_1  & b_2 & a_2+b_2 & ... & b_{r} & a_{r}+b_{r}
                                           \end{pmatrix}}.\]
Similarly, $\phi(\overline{S})=\phi _E(\left(\FF_{p^2}\right)^r)$ and the rank of each nonzero element of $\phi _E(\left(\FF_{p^2}\right)^r)$ is 2. Since the even positions of $S$ are all zeros, we can collapse the elements in such a way that the entries in the even positions are deleted and hence we can look at the elements of $\overline{S}$ as simply elements of $\left(\FF_{p^2}\right)^r$ as well.
\label{psi}
\end{remark}

\begin{theorem}
For prime $p$ where $p \equiv 2$ or 3 (mod 5) and  for any positive integer $r$, the rank-metric code $\phi_O\left(\left(\FF_{p^2}\right)^r\right)$ satisfies the Singleton bound.
\label{mrd1}
\end{theorem}

\begin{IEEEproof} 
Let $p$ be prime such that $p \equiv 2$ or 3 (mod 5), $r$ be a positive integer, and $\rho$ be the dimension of $\left(\FF_{p^2}\right)^r$ as an $\FF_p$-vector space. From Remark \ref{psi} and Theorem \ref{classtonew2}, $\phi_O(\left(\FF_{p^2}\right)^r)$ is a $[2 \times 2r, \rho, 2]$ rank-metric code. Note that $\left|\phi_O\left(\left(\FF_{p^2}\right)^r\right)\right|=p^{\rho}$ and $\left|\left(\FF_{p^2}\right)^r\right|=p^{2r}$ but $\left|\left(\FF_{p^2}\right)^r\right|=\left|\phi_O\left(\left(\FF_{p^2}\right)^r\right)\right|$. Hence it follows that $\rho = 2r$.
Now, a $[k \times \ell, \rho, \delta]$ rank-metric code satisfies the Singleton bound if \[\rho \leq \min\{k(\ell-\delta+1),\ell(k-\delta+1)\}.\]
Substituting the values,  
\[\begin{aligned}
2r & \leq \min\{2(2r-2+1),2r(2-2+1)\}\\ 2r& \leq \min\{4r-2,2r\}.
\end{aligned}\] Note that $4r-2 \ge 2r$ for $r \ge 1$.
Thus, for prime $p$ where $p \equiv 2$ or 3 (mod 5) and  for any positive integer $r$, $\phi_O\left(\left(\FF_{p^2}\right)^r\right)$ satisfies the Singleton bound for rank-metric codes and hence a maximum rank distance code.
\end{IEEEproof} 

A MAGMA program was designed to obtain the maximum rank distance code $\phi_O\left(\left(\FF_{p^2}\right)^r\right)$ for $p=2$ and for any positive integer $r$. 

In a similar manner we can prove that $\phi_E\left(\left(\FF_{p^2}\right)^r\right)$ also satisfies the Singleton bound for any positive integer $r$.

\begin{example}
Consider $\FF_4=\{0,1,\omega , 1+\omega \}$. We have \[\phi_O (\FF_4) = \left \{{\begin{pmatrix}
                                                      0 & 0  \\
                                                      0 & 0 
                                           \end{pmatrix}},
{\begin{pmatrix}
                                                      0 & 1  \\
                                                      1 & 0 
                                           \end{pmatrix}},
{\begin{pmatrix}
                                                      1 & 0  \\
                                                      1 & 1 
                                           \end{pmatrix}},
{\begin{pmatrix}
                                                      1 & 1  \\
                                                      0 & 1 
                                           \end{pmatrix}}
\right \}.\]
Note that by Theorem \ref{classtonew2} and Remark \ref{psi}, $\phi_O (\FF_4)$ is a $[2 \times 2, 2, 2]$ rank-metric code. 
By Theorem \ref{mrd1}, $\phi_O (\FF_4)$ is a maximum rank distance code.
\end{example}

\begin{example}
Again, consider $\FF_4=\{0,1,\omega , 1+\omega \}$. We have \[\phi_E (\FF_4) = \left \{{\begin{pmatrix}
                                                      0 & 0  \\
                                                      0 & 0 
                                           \end{pmatrix}},
{\begin{pmatrix}
                                                      1 & 0  \\
                                                      0 & 1 
                                           \end{pmatrix}},
{\begin{pmatrix}
                                                      0 & 1  \\
                                                      1 & 1 
                                           \end{pmatrix}},
{\begin{pmatrix}
                                                      1 & 1  \\
                                                      1 & 0 
                                           \end{pmatrix}}
\right \}.\]
Note that by Theorem \ref{classtonew2} and Remark \ref{psi}, $\phi_E (\FF_4)$ is a $[2 \times 2, 2, 2]$ rank-metric code, and a maximum rank distance code.
\end{example}

\begin{table}
\centering
\begin{tabular}{|c|c|c|}
\hline
   $\alpha$ & $\phi(\alpha)$ & $w_R(\phi(\alpha))$\\
\hline
$(0,0)$ & ${\begin{pmatrix}
                                                      0 & 0  \\
                                                      0 & 0 
                                           \end{pmatrix}}$ & 0  \\
\hline

$(0,1)$ & ${\begin{pmatrix}
                                                      0 & 1  \\
                                                      1 & 0 
                                           \end{pmatrix}}$ & 2  \\
\hline

$(0,\omega)$ & ${\begin{pmatrix}
                                                      1 & 0  \\
                                                      1 & 1 
                                           \end{pmatrix}}$ & 2  \\
\hline

$(0,1+\omega)$ & ${\begin{pmatrix}
                                                      1 & 1  \\
                                                      2 & 1 
                                           \end{pmatrix}}$ & 2  \\
\hline

$(0,2)$ & ${\begin{pmatrix}
                                                      0 & 2  \\
                                                      2 & 0 
                                           \end{pmatrix}}$ & 2  \\
\hline

$(0,2\omega)$ & ${\begin{pmatrix}
                                                      2 & 0  \\
                                                      2 & 2 
                                           \end{pmatrix}}$ & 2  \\
\hline

$(0,1+2\omega)$ & ${\begin{pmatrix}
                                                      2 & 1  \\
                                                      0 & 2 
                                           \end{pmatrix}}$ & 2  \\
\hline

$(0,2+2\omega)$ & ${\begin{pmatrix}
                                                      2 & 2  \\
                                                      1 & 2 
                                           \end{pmatrix}}$ & 2  \\
\hline

$(0,2+\omega)$ & ${\begin{pmatrix}
                                                      1 & 2  \\
                                                      0 & 1 
                                           \end{pmatrix}}$ & 2  \\
\hline

\end{tabular}
\caption{Elements of $T=\{(0,c+d\omega)|c,d \in \FF_3\}$ and their Images under $\phi$ with their Rank Weights}
\label{bachoc_rank2}
\end{table}

Now, let $T=\{(0,c+d\omega)|c,d \in \FF_3\}$. Table \ref{bachoc_rank2} shows the elements of $T$ and their corresponding images in $M_2(\FF_3)$ under $\phi$ with their rank weights. From the given table, each nonzero element of $\phi(T)$ has rank 2.

\begin{example}
Refer to Table \ref{bachoc_rank2}, $\phi(T)=\phi_O(\FF_9)$ is a $[2 \times 2, 2, 2]$ rank-metric code. 
By Theorem \ref{mrd1}, $\phi_O (\FF_9)$ is a maximum rank distance code.
\end{example}

Note that the first prime that does not satisfy Theorem \ref{mrd1} is $p=5$. We have $5 \equiv 0$ (mod 5). Now, $2+\omega \in \FF_{5^2} = \FF_{25}$ and $\phi_O (1+2\omega)= {\begin{pmatrix}
                                                      1 & 2  \\
                                                      3 & 1 
                                           \end{pmatrix}}$ whose rank is 1. Therefore, $\phi_O (\FF_{25})$ cannot be a rank-metric code with minimum distance 2.

\begin{definition}
Let $r$ and $m$ be positive integers. the matrix $H_m= {\begin{pmatrix}
                                                   I_m & 0_{m \times mr}
                                           \end{pmatrix}}$
and the matrix $\widehat H_m= {\begin{pmatrix}
                                                   0_{m \times mr} & I_m
                                           \end{pmatrix}}$, where $I_m$ is the $m \times m$ identity matrix and $0_{m \times mr}$ is the $m \times mr$ zero matrix.
\end{definition}
\vskip .10in
\begin{remark}[Anticode Bound]
\[\mathcal A_p(2r+2,4,2) \leq  \dfrac{p^{2r+2}-1}{p^2-1}\]
\label{anti2}
\end{remark}

\begin{remark}
For any natural number $r$, we have \[1+p^2+p^4+p^6+...+p^{2r}=\dfrac{p^{2r+2}-1}{p^2-1}.\]
\label{pmi}
\end{remark}

\begin{theorem}
Let $p$ be prime where $p \equiv 2$ or 3 (mod 5), $r$ be a positive integer, and consider a class of $\FF_p$-vector spaces $\left \{\left(\FF_{p^2}\right)^i| i=1,2,...,r\right \}$. Let $D_i$ be the set of vectors that contain $\Lambda \left(\phi_O \left(\FF_{p^2}\right)^i\right)$ such that the vectors are appended with zeros in the left so that they have common length $2r+2$. Then
$G_p(r,2) = \left<\widehat H_2 \right>    \bigcup \left(\displaystyle {\bigcup_{i=1}^r} D_i \right)$ is a $\left(2r+2, \dfrac{p^{2r+2}-1}{p^2-1}, 4, 2 \right)_p$ code.
\label{generaltheorem}
\end{theorem}

\begin{IEEEproof} 
For $1 \le i \le r$, let $D_i$ be the set of vectors that contain $\Lambda (\phi_O (\FF_4^i))$ such that the vectors are appended with zeros in the left so that they have common length $2r+2$.
Note that $\left|\Lambda \left(\phi_O \left( \left(\FF_{p^2}\right)^i\right)\right)\right|=\left| \phi_O \left( \left(\FF_{p^2}\right)^i\right)\right|=\left|\left(\FF_{p^2}\right)^i\right|=p^{2i}$.
Now, \[\left |\displaystyle{\bigcup_{i=1}^r} D_i \right |=p^2+p^4+p^6+...+p^{2r}.\]
\\Let $G_p(r,2) = \left<\widehat H_2 \right> \bigcup \left(\displaystyle {\bigcup_{i=1}^r} D_i \right)$ so that by Remark \ref{pmi}, \[|C|=1+p^2+p^4+p^6+...+p^{2r}=\dfrac{p^{2r+2}-1}{p^2-1}.\]
Note that the only intersection of the $D_i$'s is just the zero space. Moreover, the only intersection of $ \left<\widehat H_2 \right>$ with the $D_i$'s is also trivial.
Thus, the obtained code is a $\left(2r+2, \dfrac{p^{2r+2}-1}{p^2-1}, 4, 2 \right)_p$ code.
\end{IEEEproof} 

\begin{remark}
The code $G_p(r,2)$ obtained in Theorem \ref{generaltheorem} attains the Anticode bound given in Remark \ref{anti2}.
\end{remark}

Similarly we can also obtain a $\left(2r+2, \dfrac{p^{2r+2}-1}{p^2-1}, 4, 2 \right)_p$ code using the mapping $\phi_E$.

\begin{example}
Let $p=2$ and $r=1$. We have $\FF_4=\{0,1,\omega,1+\omega\}$. 
Now the lifted matrices of $\phi_O (\FF_4)$ are 
$ {\begin{pmatrix}
                                                   1 &  0 & 0 & 0  \\
                                                   0 &  1 &   0 & 0 
                                           \end{pmatrix}},
 {\begin{pmatrix}
                                                   1 &  0 &   0 & 1  \\
                                                    0 &  1 &  1 & 0 
                                           \end{pmatrix}},
 {\begin{pmatrix}
                                                    1 &  0 &  1 & 0  \\
                                                    0 &  1 &  1 & 1 
                                           \end{pmatrix}},
 {\text {and } } $\\  $ {\begin{pmatrix}
                                                   1 &  0 &   1 & 1  \\
                                                   0 &  1 &   0 & 1 
                                           \end{pmatrix}}.$
Then the elements of $G_2(1,2)$ are
 \[\begin{aligned}
C_1 & =\{(1,0,0,0),(0,1,0,0),(1,1,0,0),(0,0,0,0)\},\\
C_2 & =\{(1,0,0,1),(0,1,1,0),(1,1,1,1),(0,0,0,0)\},\\
C_3 & =\{(1,0,1,0),(0,1,1,1),(1,1,0,1),(0,0,0,0)\},\\
C_4 & =\{(1,0,1,1),(0,1,0,1),(1,1,1,0),(0,0,0,0)\}, {\text { and} };\\
C_5 & =\{(0,0,1,0),(0,0,0,1),(0,0,1,1),(0,0,0,0)\}.
\end{aligned}\]
Note that $G_2(1,2)$ is a $(4,5,4,2)_2$ code. 
Now, when $p=2$ and $r=1$, the Anticode bound becomes $\dfrac{2^{2+2}-1}{3}=5$. Thus, $G_2(1,2)$ attains this bound.
\label{exanticode}
\end{example}

\begin{example}
Again, consider $\FF_4=\{0,1,\omega,1+\omega\}$. 
Now the lifted matrices of $\phi_E (\FF_4)$ are 
$ {\begin{pmatrix}
                                                   1 &  0 & 0 & 0  \\
                                                   0 &  1 &   0 & 0 
                                           \end{pmatrix}},
 {\begin{pmatrix}
                                                   1 &  0 &   1 & 0  \\
                                                    0 &  1 &  0 & 1 
                                           \end{pmatrix}},
 {\begin{pmatrix}
                                                    1 &  0 &  0 & 1  \\
                                                    0 &  1 &  1 & 1 
                                           \end{pmatrix}},
 {\text {and } } $\\  $ {\begin{pmatrix}
                                                   1 &  0 &   1 & 1  \\
                                                   0 &  1 &   1 & 0 
                                           \end{pmatrix}}.$
Then the elements of the Grassmannian code $\mathcal C$ generated by the lifted matrices, with  $C_5=\left<\widehat H_2 \right>$ are given by
 \[\begin{aligned}
C_1 & =\{(1,0,0,0),(0,1,0,0),(1,1,0,0),(0,0,0,0)\},\\
C_2 & =\{(1,0,1,0),(0,1,0,1),(1,1,1,1),(0,0,0,0)\},\\
C_3 & =\{(1,0,0,1),(0,1,1,1),(1,1,1,0),(0,0,0,0)\},\\
C_4 & =\{(1,0,1,1),(0,1,1,0),(1,1,0,1),(0,0,0,0)\}, {\text { and} };\\
C_5 & =\{(0,0,1,0),(0,0,0,1),(0,0,1,1),(0,0,0,0)\}.
\end{aligned}\]
Note that $\mathcal C$ is also a $(4,5,4,2)_2$ code that attains the Anticode bound.
\label{exanticode}
\end{example}

\begin{corollary}
The dual of $G_p(r,2)$ is a $\left(2r+2, \dfrac{p^{2r+2}-1}{p^2-1}, 4, 2r \right)_p$ code.
\label{generaltheorem2}
\end{corollary}

\begin{IEEEproof} 
This directly follows from Theorem \ref{ortho}.
\end{IEEEproof} 

MAGMA programs were created to obtain the anticode-optimal code $G_p(r,2)$ and its dual in Theorem \ref{generaltheorem} and Corollary \ref{generaltheorem2} for $p=2$ and for any positive integer $r$.

\section{Graphs of Anticode-Optimal Grassmannian Codes $G_p(r,2)$}
\label{complete:graphs}

A {\it graph} $G$ is a pair $(V,E)$ where $V$ is a finite set whose members are called {\it vertices}, and $E$ is a subset of the set $V \times V$ of unordered pairs of vertices. The members of $E$ are called {\it edges} \cite{bigwhit}. If $\{v,w\}$ is an edge of $G$, the vertices $v$ and $w$ are said to be {\it adjacent}. An edge with identical ends is called a {\it loop} and an edge with distinct ends is called a {\it link}. A graph is {\it simple} if it has no loops and no two of its links join the same pair of vertices. In a simple graph, the {\it degree} of a vertex $v \in G$ is the number of edges of $G$ incident with $v$ {\cite{bondy}}.

A simple graph in which each pair of distinct vertices is joined by an edge is called a {\it complete graph}. A complete graph with $N$ vertices is denoted by $K_N$. The complete graph of $N$ vertices has $\dfrac{N(N-1)}{2}$ edges. The degree of any vertex in $K_N$ is $N-1$.

Note that for distinct $A, B \in G_p(r,2)$ in Theorem \ref{generaltheorem}, we have $A \cap B =\{0\}$.
\begin{definition}
Let the subspaces of $G_p(r,2)$ be the vertices of the graph $\Gamma_p(r,2)$. Two vertices $A$ and $B$ are adjacent if and only if $\dim(A\cap B)=0$.
\end{definition}

It follows that the edge set of $\Gamma_p(r,2)$ is the set of all unordered distinct pair of vertices.
\begin{theorem}
The graph $\Gamma_p(r,2)$ is a complete graph with $\dfrac{p^{2r+2}-1}{p^2-1}$ vertices.
\label{grassmanntree}
\end{theorem}

\begin{IEEEproof} Note that $|G_p(r,2)|=\dfrac{p^{2r+2}-1}{p^2-1}$ so $\Gamma_p(r,2)$ has $\dfrac{p^{2r+2}-1}{p^2-1}$ vertices. Since the intersection of any two subspaces in $G_p(r,2)$ is trivial, its dimension is zero. Thus, each pair of vertices is joined by an edge. By definition, $\Gamma_p(r,2)$ is a complete graph with $\dfrac{p^{2r+2}-1}{p^2-1}$ vertices. \end{IEEEproof} 

\begin{remark}
We can easily compute the number of edges of $\Gamma_p(r,2)$ and the degree of each vertex.
\end{remark}

\begin{example}
When $p=2$ and $r=2$, we have a $(4,21,4,2)_2$ code. The associated graph $\Gamma_2(2,2)$ of the $(4,21,4,2)_2$ code is a complete graph with 21 vertices. The number of edges is 210 and the degree of each vertex is 20.
\end{example}


\begin{thebibliography}{DGS}
\bibitem{ahlswede}
R. Ahlswede and N. Cai and S.-Y. Li and R. Yeung, ``Network information flow'', {\textsl{IEEE Trans. Inf. Theory}}, vol. 46, no. 4, pp. 1204-1216, 2000.

\bibitem{bachoc}
C. Bachoc, ``Application of coding theory to the construction of modular lattices," \textsl{J. Combinatorial Theory}, vol. 78, pp. 92-119, 1997.

\bibitem{bigwhit}
N.~L. Biggs and A.~T. White, Permutation Groups and Combinatorial
  Structures, Cambridge University Press, New York, 1979.

\bibitem{bondy}
J. A. Bondy and U. S. R. Murty, Graph Theory With Applications, \textsl{Elsevier Science Publishing Co., Inc.}, New York, 1976.

\bibitem{etz}
T. Etzion, ``Subspace codes $-$ bounds and constructions'', {\textsl{1st European Training School on Network Coding, Bacelona, Spain}}, February 2013.

\bibitem{etzion:vardy}
T. Etzion and  A. Vardy, Error-correcting codes in projective space, {\textsl{IEEE Trans. Inf. Theory}}, vol. 57, no. 2, pp. 1165-1173, February 2011.

\bibitem{falcunit:sison}
D. Falcunit, Jr. and V. Sison, “Cyclic Codes over the Matrix Ring $M_2(\FF_p)$ and their Isometric Images
over $\FF_{p^2} + u\FF_{p^2}$ ”, {\textsl{Proceedings of the 2014 International Z$\ddot{\text{u}}$rich Seminar on Communications, Sorell Hotel Z$\ddot{\text{u}}$richberg, Z$\ddot{\text{u}}$rich, Switzerland}}, pp. 91-96, 26-28 February 2014.

\bibitem{khaleghi}
A. Khaleghi and D. Silva and F. R. Kschischang, Subspace Codes, \textsl{IMA Int. Conf.}, vol. 49, no. 4, pp. 1-21, 2009.

\bibitem{kk}
R. K\"otter and F. R. Kschichang, Coding for errors and erasures in random network coding, \textsl{IEEE Trans. Inf. Theory}, vol. 54, no. 8, pp. 3579-3591, 2008.

\end{thebibliography}
\end{document}